\documentclass[prl,a4paper,superscriptaddress,twocolumn,showpacs,amsmath,amssymb,floatfix,amstex]{revtex4}

\usepackage{graphicx}

\newcommand{\Pch}{P_\text{ch}}
\newcommand{\Wtr}{W_\text{tr}}
\newcommand{\energy}{E}
\newcommand{\DT}{\Delta T}
\newcommand{\e}{\mathrm{e}}

\newcommand{\mysection}[1]{\section*{#1}}

\newcommand{\leadparagraph}[1]{\textbf{#1}}

\begin{document}

\title{Scaling of Chaos in Strongly Nonlinear Lattices}

\author{Mario Mulansky} 
\email{mulansky@pks.mpg.de}
\affiliation{\mbox{Max-Planck-Institut f\"ur Physik komplexer Systeme, N\"othnitzer Str.\ 38, D-01187 Dresden, Germany}}
\affiliation{\mbox{Institute for Theoretical Physics, TU Dresden, Zellescher Weg 17, D-01069 Dresden, Germany}}
\affiliation{\mbox{Department of Physics and Astronomy, Potsdam University, 
  Karl-Liebknecht-Str 24, D-14476, Potsdam-Golm, Germany}}
\date{\today}

\begin{abstract}
\noindent Although it is now understood that chaos in complex classical systems is the foundation of thermodynamic behavior, the detailed relations between the microscopic properties of the chaotic dynamics and the macroscopic thermodynamic observations still remain mostly in the dark.
In this work, we numerically analyze the probability of chaos in strongly nonlinear Hamiltonian systems and find different scaling properties depending on the nonlinear structure of the model.
We argue that these different scaling laws of chaos have definite consequences for the macroscopic diffusive behavior, as chaos is the microscopic mechanism of diffusion.
This is compared with previous results on chaotic diffusion [New J.\ Phys.\ 15, 053015 (2013)], and a relation between microscopic chaos and macroscopic diffusion is established.
\end{abstract}


\maketitle

\leadparagraph{The relation of the properties of microscopic chaos with the macroscopic phenomenon of diffusion has been studied since Chirikov's early works in the late 70s~\cite{Chirikov_79}.
Recently, subdiffusive spreading in disordered nonlinear systems was found and studied extensively.
Analyzing the chaotic behavior as the origin of this subdiffusive process will give further insight in the properties of the spreading and will help to explain some controversially discussed observations.
Quantifying the probability of chaos in such systems will lead to some qualitative predictions for the spreading depending on the structure of the nonlinearity.
However, a full quantitative relation of the microscopic dynamics and the macroscopic diffusion will remain out of reach.
}

\mysection{Introduction}

Bridging the gap between chaotic trajectories in the microscopic perspective and the macroscopic phenomenology of thermodynamic behavior is one of the fundamental problems of statistical physics and still remains an open challenge for theoreticians.
In recent years, substantial progress has been made in this direction by studying the diffusive behavior in nonlinear disordered models.
The most prominent observation in this respect is the spreading of energy in systems with disorder when adding nonlinear interactions.
In the absence of nonlinearity, the disorder in these models induces localization and thus the ``Absence of diffusion''~\cite{Anderson_58}, later called Anderson localization~\cite{Lee_Ramakrishnan:85}.
In recent studies it was found that an additional nonlinearity seems to re-establish diffusive properties as initially localized excitations exhibit energy spreading in such situations~\cite{Pikovsky_Shepelyansky_08,Mulansky_Pikovsky_10,Flach_Krimer_Skokos_09,laptyeva2010crossover,Fishman-Krivolapov-Soffer-12}.

This phenomenon was first understood as the ``destruction of Anderson localization by weak nonlinearities''~\cite{Pikovsky_Shepelyansky_08} and studied extensively in numerical experiments using the Discrete Nonlinear Schr\"odinger Equation with local disorder, e.g.~\cite{Pikovsky_Shepelyansky_08,Mulansky_Pikovsky_10,Flach_Krimer_Skokos_09,laptyeva2010crossover}, and~\cite{Fishman-Krivolapov-Soffer-12} for a recent review.
Later, it was realized that this effect can be viewed in the broader context of ``chaotic diffusion'' which led to an increased understanding of the spreading phenomenon~\cite{Mulansky_Pikovsky_13,Mulansky_phd}.
The most prominent observation is a subdiffusive spreading of the width of an initially localized wave packet~$L\sim t^\nu$, $\nu<1/2$~\cite{Pikovsky_Shepelyansky_08,Mulansky_Pikovsky_10,Flach_Krimer_Skokos_09}.
In contrast to these numerical results, rigorous arguments showed that on very long time scales spreading has to be slower than any power law~\cite{Wang_Zhang_09}, while a perturbation analysis of the nonlinearity suggested a propagating front moving as $\ln t$ beyond which localization occurs~\cite{Fishman-Krivolapov-Soffer-09}.
It is, however, unclear if those analytical results are applicable for the parameter regimes and time scales considered in the numerical experiments.
Nevertheless, there also exist numerical results indicating a slowing down of spreading away from the subdiffusive power-law, based on the scaling of chaos~\cite{Pikovsky_Fishman_11} but also on direct numerical observations of spreading in nonlinear Hamiltonian oscillator chains~\cite{Mulansky_Ahnert_Pikovsky_11,Mulansky_Pikovsky_13}.
A possible mechanism for the devation from power-law spreading was presented in~\cite{Michaely_12}.

Here, we want to draw further connection between the microscopic chaos and the diffusion properties in strongly nonlinear Hamiltonian chains introduced in earlier works~\cite{Mulansky_Ahnert_Pikovsky_11}.
Specifically, we look at the probability of chaos in chains of nonlinear oscillators, which is the probability to have a chaotic trajectory when starting from some random initial condition, thus it is also a measure for the fraction of the phase space belonging to the chaotic component.
We analyze the scaling of this probability around the transition point from chaotic to regular phase space in dependence of the system size.
We find in numerical studies that this scaling depends crucially on the choice of nonlinearity.
By connecting the scaling behavior of chaos with the phenomenon of chaotic diffusion in such systems we are able to deduce the existence of fundamentally different spreading laws.
Indeed, such a difference in the spreading behavior has been observed in nonlinear oscillator chains~\cite{Mulansky_Ahnert_Pikovsky_11,Mulansky_phd} indicating that the detailed properties of chaos play a fundamental role for thermodynamic effects.

This article is structured as follows: first we introduce the model of Hamiltonian oscillator chains, followed by a description of the numerical procedure to measure the probability of chaos.
Then we present our results and analyze the scaling of chaos and finally the implications for nonlinear spreading are discussed.

\mysection{Strongly Nonlinear Lattices}

The subject of study in this work are Hamiltonian chains of harmonic or nonlinear oscillators with nonlinear nearest-neighbor coupling.
Such models are called strongly nonlinear because of the absence of linear coupling terms and hence the absence of linear waves.
They belong to the class of systems considered in~\cite{Froehlich_Spencer_Wayne:86}.
The Hamilton function for this model is written in terms of position~$q_k$ and momentum~$p_k$ of the oscillator at site~$k$:
\begin{equation} \label{eqn:hamilton_function}
 H = \sum_{k=1}^{N}\left( \frac{p_k^2}{2} + \frac{\omega_k^2}{\kappa} q_k^\kappa \right) + \frac{1}\lambda \,\sum_{k=1}^{N-1} (q_{k+1} - q_k)^\lambda\;.
\end{equation}
The on-site potential has power $\kappa\geq2$, so we consider harmonic or nonlinear local oscillators, while the coupling is strictly nonlinear $\lambda>2$ and of higher power $\lambda>\kappa$.
In the following, we will consider two choices of the nonlinear powers: first a fully nonlinear model where $\kappa=4$ and $\lambda=6$, called \textbf{model A}.
The second choice are harmonic oscillators with nonlinear coupling: $\kappa=2$, $\lambda=4$ which will be called \textbf{model B}.
Furthermore, we consider regular lattices with $\omega_k=1$, but also the case of disorder where $\omega_k\in[1/2,3/2]$ chosen uniformly and independently identically distributed.
For the fully nonlinear case in Model A, the introduction of disorder makes no fundamental difference for the scaling of chaos as will be seen later.
For harmonic oscillators, on the other hand, disorder in terms of random oscillator frequencies is crucial as otherwise the whole chain would always be in resonance which strongly favors the existence of chaos.
Note, that for studying chaotic diffusion one usually considers disordered systems to eliminate nonlinear waves~\cite{Mulansky_Pikovsky_13}.
For fixed nonlinearities $\kappa$ and $\lambda$ and disorder~$\omega_k$, the only parameter in model~\eqref{eqn:hamilton_function} is the conserved total energy $E=H(q,p)$.
An important observable in this system is the local energy at site~$k$: 
\begin{equation} \label{eqn:local_energy}
E_k = \frac{p_k^2}{2} + \frac{\omega_k^2}{\kappa} q_k^\kappa + \frac1{2\lambda} [(q_{k+1} - q_k)^\lambda + (q_{k} - q_{k-1})^\lambda].
\end{equation}
From $\lambda>\kappa$ it then follows that in the limit of small local energies $E_k\rightarrow0$ the system~\eqref{eqn:hamilton_function} becomes an integrable chain of uncoupled oscillators.
Assuming a roughly uniform energy distribution, this corresponds to the limit of vanishing energy density~$W\rightarrow0$, where $W = \langle E_k\rangle = E/N$.
Consequently, for $W\rightarrow0$ one expects a completely regular phase space without chaotic regions.

Note, that system~\eqref{eqn:hamilton_function} can also be parametrized differently by introducing a coupling parameter $\beta$:
\begin{equation} \label{eqn:scaled_hamilton}
 H' = \sum_{k=1}^{N}\left( \frac{p_k'^2}{2} + \frac{\omega_k^2}{\kappa} q_k'^\kappa \right)+ \frac{\beta}\lambda \,\sum_{k=1}^{N-1} (q_{k+1}' - q_k')^\lambda\;.
\end{equation}
This involves rescaled amplitudes $q_k'$, $p_k'$ and a rescaled time $t'$ and gives the coupling strength~$\beta$ as the only parameter while the dynamics become independent from the total energy~$E'$~\cite{Mulansky_phd}.
The transformation is given by:
\begin{equation} \label{eqn:nonhomogeneous_transformation}
\begin{aligned}
q_k &= \beta^{b} q_k', &
p_k &= \beta^{\kappa b/2} p_k' \\
t &= \beta^{(2-\kappa)b/2} t', &
H &= \beta^{\kappa b} H'
\end{aligned}
\end{equation}
with $b = 1/(\lambda-\kappa)$.
Fixing the rescaled total energy to $E'=N$, one finds that changing the nonlinear strength~$\beta$ in~\eqref{eqn:scaled_hamilton} corresponds to a change of the original energy~$E$ in~\eqref{eqn:hamilton_function} according to:
\begin{equation} \label{eqn:energy_scaling}
 E=N\beta^{\kappa/(\lambda-\kappa)}.
\end{equation}
For the numerical simulations presented below, we used this second parametrization~\eqref{eqn:scaled_hamilton}.
This is especially valuable for simulations of model A, where $\kappa=4$ and $\lambda=6$, as there for formulation~\eqref{eqn:hamilton_function} the uncoupled oscillator frequencies and thus the fundamental time-scales depend on the nonlinearity parameter~$E$.
This would imply that the simulation times need to be increased for smaller densities $W=E/N$, but accordingly the simulation time-step could be increased as also the dynamics slows down.
In the rescaled formulation however, the energy density and thus the typical frequencies are kept constant and only the nonlinear coupling parameter~$\beta$ is changed.
Hence the parametrization in~\eqref{eqn:scaled_hamilton} is better suited for numerical simulations as it allows a fixed integration time and time step for all values of the nonlinearity~$\beta$, while for the original model~\eqref{eqn:hamilton_function} it is easier to connect the results to the spreading behavior of trajectories.

\mysection{Probability of Chaos}

We will now examine the emergence of regular regions in phase space when approaching the integrable limit of zero density.
Therefore, a measure of the fraction of the chaotic phase space is introduced which we call the ``probability of chaos'' $P_\text{ch}$.
This quantity has already been followed for the Discrete Nonlinear Schr\"odinger Equation~\cite{Pikovsky_Fishman_11} and Hamiltonian oscillator chains~\cite{Mulansky_Ahnert_Pikovsky_Shepelyansky_11,basko2012local}.
$P_\text{ch}$ is measured numerically by categorizing trajectories as being chaotic or regular based on the largest Lyapunov exponent~$\lambda$.
Therefore, we simulate a trajectory according to~\eqref{eqn:hamilton_function} starting from a \emph{random} initial condition in equilibrium (uniform energy distribution) and calculate~$\lambda$ by standard methods.
As explained above, we use the parametrization~\eqref{eqn:scaled_hamilton} for our numerical simulations.
The initial conditions are chosen in the following way: first the positions~$q_k$ and momenta~$p_k$ of the oscillators are initialized as random numbers  chosen from a normal distribution with zeros mean and variance $\sigma^2$: $\mathcal{N}(0,\sigma)$.
The variance is calculated such that the expectation value of the energy density has the desired value $\langle E'_k \rangle = 1$, i.e.\ $E'=N$, where the average here is taken over many initial conditions.
Then, in a correction step, the momenta are scaled such that the actual energy density of this initial condition has exactly the desired value $E'(p,q)/N=1$.
The numerical time evolution of the trajectories is computed using a fourth order symplectic algorithm~\cite{McLachlan_95} and we performed the simulations on GPU devices using the Boost.odeint C++ library~\cite{ahnert2011odeint,demidov2013programming}.

\begin{figure}[t]
  \centering
  \includegraphics[width=0.48\textwidth,angle=0]{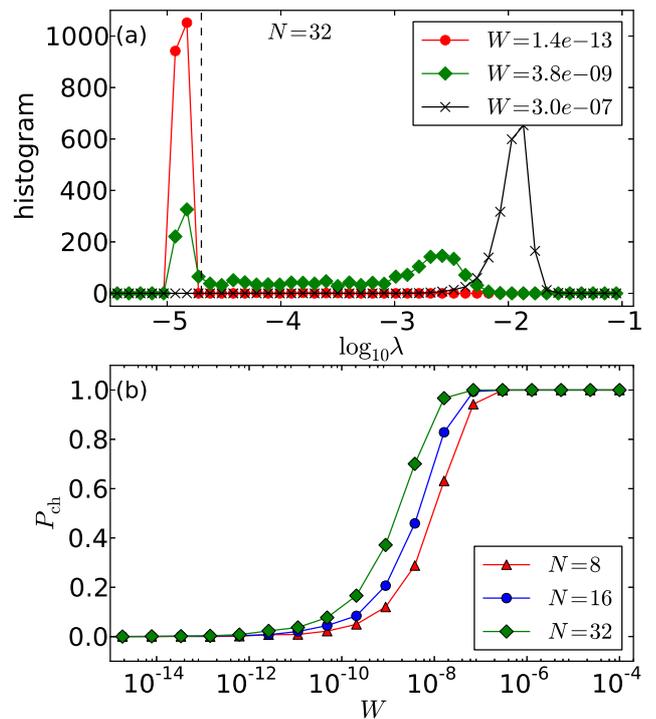}
  \caption{(color online) Histogram of the Lyapunov exponent $\lambda$ for different energy densities~$W$ in Model A with $N=32$ sites (panel a). Panel (b) shows the fraction of chaotic trajectories $P_\text{ch}$, where ${\lambda>2\cdot10^{-5}}$, dashed line in (a), in dependence of the energy density for several system sizes $N=8,16,32$.}
\label{fig:lyap_prob}
\end{figure}

We compute the Lyapunov exponents for ${M=2000}$ such initial conditions for several lattice sizes~$N$ and coupling parameter~$\beta$.
Then we perform a back-transformation to parametrization~\eqref{eqn:hamilton_function} with ${W=\beta^{\kappa/(\lambda-\kappa)}}$ according to~\eqref{eqn:energy_scaling}.
The result is a distribution of $\lambda$ for each parameter value $N$ and $W$.
Figure~\ref{fig:lyap_prob}a shows histograms of these Lyapunov exponents for Model A of length $N=32$ with a regular local potential~$\omega_k=1$ and energy densities ${W=1.4\cdot10^{-13},}\, {3.8\cdot10^{-9},}\, {3\cdot10^{-7}}$.
One clearly sees that for small densities almost all Lyapunov exponents are close to $\lambda\approx10^{-5}$.
For regular dynamics the maximum Lyapunov exponent should be exactly zero, but numerically, with finite integration time $T=10^6$, such small values are below the numerical accuracy and essentially indicate regular motion.
For larger energy densities one finds Lyapunov exponents of the order of $\lambda\approx 0.01$, clearly indicating chaotic dynamics.

Based on this observation we introduce a threshold value $\lambda_c=20/T=2\cdot10^{-5}$ and consider all trajectories with smaller Lyapunov exponents $\lambda<\lambda_c$ as regular, and those with larger values chaotic $\lambda>\lambda_c$.
Note, that this threshold is solely defined by the finite integration time.
Longer integrations would allow for smaller $\lambda_c$.
Using this distinction of regular and chaotic trajectories we can straight forwardly define the probability of chaos as the fraction of chaotic trajectories: $P_\text{ch}=M_\text{ch}/M$, where $M_\text{ch}$ is the number of chaotic trajectories and $M=2000$ the total number of trajectories.
Figure~\ref{fig:lyap_prob}b shows $P_\text{ch}(W)$ for the regular Model A and several chain lengths ${N=8,\,16,\,32}$.
One can clearly see the sigmoidal transition with the limits $\Pch\rightarrow0$ for $W\rightarrow0$, the integrable limit, and $\Pch\rightarrow1$ for $W\rightarrow\infty$, the strongly nonlinear limit.
While in the limits of small and large densities one finds clear peaks in the Lyapunov distribution (red circles and black crosses in Figure~\ref{fig:lyap_prob}a), in the intermediate regime at the transition a plateau of intermediate Lyapunov values $\lambda=10^{-5}\dots10^{-3}$ emerges (green diamonds in Figure~\ref{fig:lyap_prob}a).
These intermediate Lyapunov values probably belong to weakly chaotic trajectories~\cite{Mulansky_Ahnert_Pikovsky_Shepelyansky_11} where the dynamics are chaotic, but on long time scales influenced by the regular islands in phase space.
However, it can not be excluded at this point that these values have not yet converged and might for a longer simulation approach zero or some larger finite value.
Nonetheless, we have verified that the fundamental observation, the scaling of chaos presented below, is independent from the exact choice of $\lambda_c$ and thus also from those weakly chaotic or non-converged cases.

\mysection{Scaling}

We now focus on the size dependence of the transition point $\Wtr(N)$ defined by $\Pch(\Wtr,N)=1/2$.
It is known that chaos emerges from resonances~\cite{Chirikov_79} and thus it is clear that for larger systems the transition point~$\Wtr$ should decrease because with more oscillators one also has more possibilities of resonances.
This can be seen in Figure~\ref{fig:lyap_prob}b already.
An intuitive assumption for such oscillator chains is the \emph{locality of chaos}, which has been verified previously~\cite{Pikovsky_Fishman_11,basko2011weak,basko2012local}.
This means essentially that if two chains are connected then the \emph{probability of regular motion} in the combined system is given by the product of probabilities of the two constituents.
The underlying idea is that the combined system is regular only if both subsystems are regular, and that the probabilities of the subsystems are independent.
From this assumption one finds that the probability of chaos can be written as $\Pch = 1-\e^{-g(W) N}$ with some function $g(W)$ independent of $N$~\cite{basko2011weak}.
Here, we are interest in the relation of the transition point $\Wtr$ and the system size $N$, but therefore we need to know the function $g(W)$ as well.
For nonlinearly coupled harmonic oscillators with random frequencies (model B) Basko calculated $g(W)$ in the asymptotic limit $W\rightarrow0$ by assuming that chaos is created from double resonances, which gives quite naturally $g(W\rightarrow0)\sim W^2$~\cite{basko2012local}.
The transition point $\Wtr$, however, is far from this asymptotic regime of small densities, and in a numerical study Basko found $g(W)\sim W^\beta$ with $\beta\approx2.85\pm0.1$ for intermediate densities~\cite{basko2012local}.
Thus, we will also assume a simple power-law dependence $g(W)\sim W^\beta$ here and deduce the exponent from numerical simulations.
The power law dependence means the chaos probability $\Pch$ should become independent of the system size~$N$ when using the scaled coordinates $N^\alpha W$, where $\alpha=1/\beta$, i.e.\ $\Pch(W,N) = f(N^\alpha W)$.

\begin{figure}[t]
  \centering
  \includegraphics[width=0.48\textwidth,angle=0]{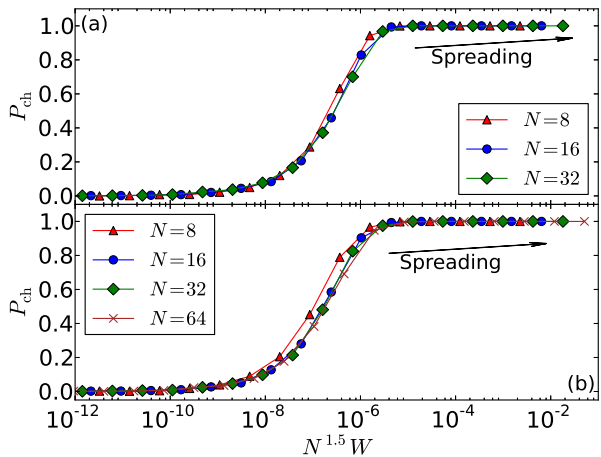}
  \caption{(color online) Probability of chaos for Model A without disorder $\omega_k=1$ (a -- same data as in Figure~\ref{fig:lyap_prob}b) and with disorder $\omega_k\in[1/2,3/2]$ (b) in dependence of the scaled density $N^{1.5} W$.}
\label{fig:4_6}
\end{figure}

For the case of the fully nonlinear systems with $\kappa=4$ and $\lambda=6$ we found that an exponent of $\alpha_{46}=1.5$ corresponds well to our results.
This is shown in Figure~\ref{fig:4_6} for a regular chain ($\omega_k=1$, Figure~\ref{fig:4_6}a) and the disordered case ($\omega_k\in[1/2,3/2]$, Figure~\ref{fig:4_6}b).
There, we plot the probability of chaos $\Pch$ as function of the rescaled energy density $N^{1.5}\,W$ and find a perfect overlap of the curves for different sizes $N=8\dots64$.
This shows that increasing the system size indeed only shifts the transition to smaller densities, but does not change the shape of the transition, which verifies our approach.
We note that the introduction of local disorder does not change the behavior of $\Pch$.
Again, this scaling is only valid for rather large densities around the transition point.
We do not expect it to represent the asymptotic behavior for vanishing densities.
The asymptotic limit $\Pch(W\rightarrow0)$ is not addressed here, but we briefly note that the $W^2$ behavior derived for harmonic oscillators does not hold for nonlinear oscillators, because due to the amplitude dependent frequencies the resonance structure of nonlinear oscillators is completely different from the harmonic case.

\begin{figure}[t]
  \centering
  \includegraphics[width=0.48\textwidth,angle=0]{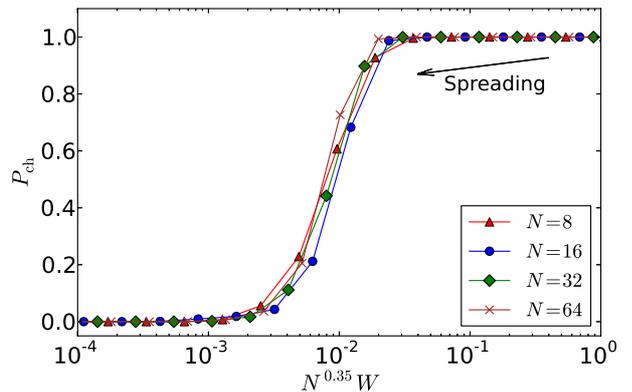}
  \caption{(color online) Probability of chaos for Model B with local disorder $\omega_k\in[1/2,3/2]$ in dependence of the scaled density $N^{0.35} W$.}
\label{fig:2_4}
\end{figure}

For the case of harmonic oscillators with random frequencies $\omega_k\in[1/2,3/2]$ and nonlinear coupling, $\kappa=2$, $\lambda=4$, the scaling exponent was found to be $\alpha_{24}=0.35$.
Notably, this value is in perfect agreement with the numerical results reported by Basko, who found for the intermediate regime that $g(W)\sim W^{2.85}$~\cite{basko2012local}, which translates to ${\alpha=1/2.85=0.35}$.
The numerical results with a rescaled density according to this parameter value are shown in Figure~\ref{fig:2_4}.
Again, the rescaled curves show a convincing overlap indicating also here that changing the system size only shifts the transition point.
So we find that for both cases, nonlinear and harmonic local oscillators, the size dependence of the transition from chaotic to regular phase space can be modeled with a simple power-law $\Wtr \sim N^{-\alpha}$.
The numerical value for the exponent, however, is significantly different: for the fully nonlinear case we find $\alpha_{46}=1.5$, while for harmonic oscillators we get $\alpha_{24}=0.35$.
This different scaling has its origin in the different resonance structure of harmonic and nonlinear oscillators, but a rigorous derivation of the exact scaling exponents~$\alpha$ is beyond the scope of this work.

It should be noted, that we have verified that the scaling of chaos presented above is absolutely robust against the details of the simulation and our definition of chaos.
E.g.\ changing the threshold for chaos to $\lambda_c=10^{-4}$ or $\lambda_c=10^{-3}$ broadens the transition from chaos to regularity, but gives exactly the same scaling behavior.
Also, we have verified that using an integration time of $T=10^{5}$ gives the same scaling observation, and also test runs with $T=10^7$ did not show any difference.

\mysection{Implications for Spreading}

The scaling law identified above has some remarkable consequences for the spreading process in such setups.
This spreading is typically observed in the following sense: One starts with only a few excited oscillators in the center of a large chain of model~\eqref{eqn:hamilton_function} while all other oscillators are at rest. Then, as time evolves, more and more oscillators become excited due to the nearest-neighbor interactions.
The number of excited oscillators, usually called excitation width~$L$, grows but consequently the energy density of the excited oscillators decreases because the conserved total energy gets distributed over more sites.
The relation to the scaling of chaos described above can be found from assuming that the phase space of the short chains with length~$N$ investigated here also effectively models the phase space around a spreading trajectory currently extended over $L=N$ sites.
Then, increasing the number of sites can, in this sense, be interpreted as an \emph{effective increase of the dimensionality} of phase space accessed by the trajectory.
So assuming $W=\energy/L$ and $N\sim L$, the scaled variable in Fig.~\ref{fig:4_6} \emph{increases} for spreading states as ${N^{1.5}W \sim L^{1.5}W \sim L^{0.5}\energy}$.
That means in the course of spreading the trajectory is driven \emph{away} from the regular parts of the phase space!
This is a quite surprising, even counter-intuitive result as one would naturally expect that for small energy densities eventually a KAM-regime will be reached.
But this is only true for a fixed system size.
For spreading states, however, together with decreasing~$W$, also the effective dimensionality is increased and it turns out numerically that for model A the latter effect is stronger ($\sim N^{1.5}$ vs.\ $\sim 1/N$).
This is also indicated by the arrows in Figure~\ref{fig:4_6}.
Hence, the KAM regime will not be reached for spreading states in this case.
This is true for both regular and disordered potential in model A with $\kappa=4$, $\lambda=6$.

For the harmonic oscillators in model B, the scaling of chaos was found above as $N^{0.35}$.
With the same assumptions as before this means that the probability of chaos for spreading states \emph{decreases} as $N^{0.35}W\sim E/L^{0.65}$ with time.
The trajectory thus moves towards the regular KAM-regime while spreading, where more and more regular islands appear in the phase space.
Hence the typical phase space environment for spreading trajectories is fundamentally different for nonlinear (model A, $\kappa=4$) and harmonic (model B, $\kappa=2$) oscillators.
Therefore, one can expect a qualitative difference in the spreading behavior for these two cases.
It should be noted, however, that a quantitative implication from the different scaling observation for the spreading can not be deduced.
Moreover, the fact that the spreading trajectories are driven towards the regular regime does not necessarily mean that the spreading process has to stop at some point.
Although the phase space becomes more and more regular there always exist thin layers of chaos where the trajectory can travel along, a mechanism called Arnol'd Diffusion~\cite{arnold1964instability}.

\begin{figure}[t]
  \centering
  \includegraphics[width=0.42\textwidth,angle=0]{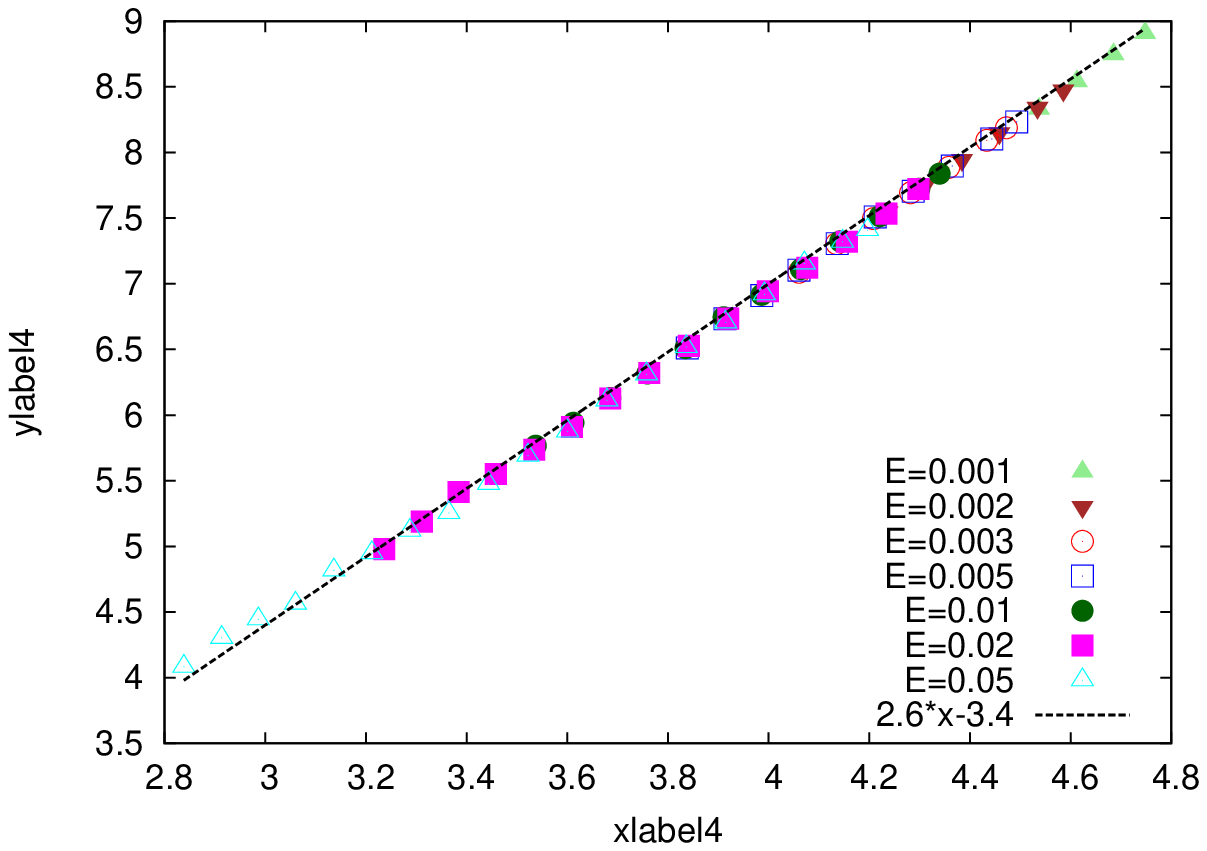}\\[.5cm]
  \includegraphics[width=0.4\textwidth,angle=0]{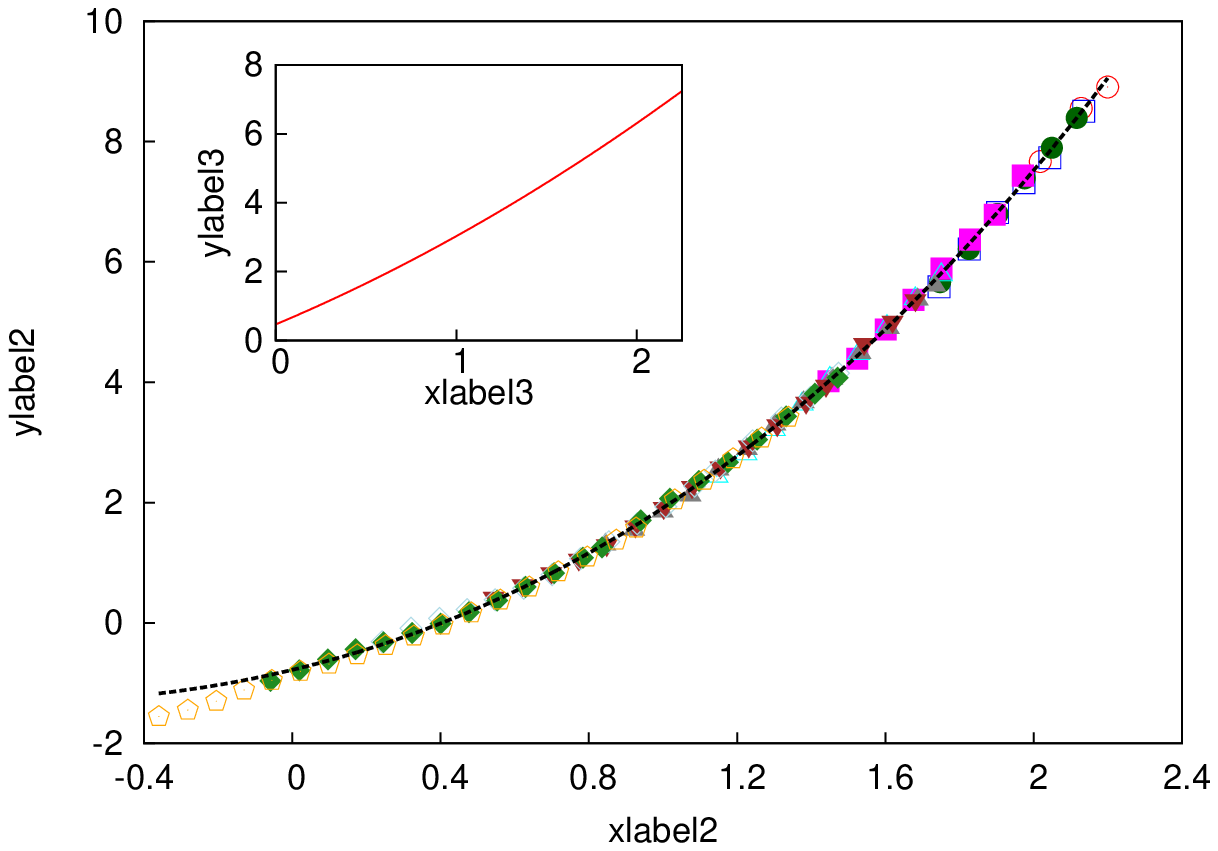}
  \caption{(color online) Spreading behavior for Model A (upper graph) and Model B (lower graph) in terms of the excitation times~$\Delta T$ as function of the excitation length~$L$. Plot reproduced from New J.\ Phys.\ 15, 053015, ``Energy spreading in strongly nonlinear disordered lattices'' (2013)~\cite{Mulansky_Pikovsky_13}. The inset shows the slope of the spreading curve $a(W)$ in model B.}
\label{fig:spreading}
\end{figure}

Nevertheless, a qualitative difference for spreading in fully nonlinear and harmonic oscillators has indeed been reported recently~\cite{Mulansky_Ahnert_Pikovsky_11,Mulansky_Pikovsky_13}.
There, the spreading process was quantified in terms of the average time required to excite one new oscillator~$\DT$ as function of the current energy density $W=E/L$.
For details on this observable, its averaging and scaling with the energy~$E$ we refer to~\cite{Mulansky_Pikovsky_13}.
Here, we only want to emphasize the fact that for model A a pure power-law spreading was found, while model B showed a clear deviation from the power-law behavior and thus a qualitatively different spreading.
This is illustrated in Figure~\ref{fig:spreading} showing some results already reported in~\cite{Mulansky_Pikovsky_13}.
The crucial observation in these results is the pure power law behavior $\DT\sim W^a$ for model A (upper panel in Figure~\ref{fig:spreading}), while model B (lower panel in Figure~\ref{fig:spreading}) shows a clear deviation from such a power law.
This is further emphasized in the inset where the numerical slope of the spreading curve for model B, ${a(W) = \frac{\textrm{d}\log_{10}\Delta T/L}{\textrm{d}\log_{10}W}}$, is plotted.
The deviation from a pure power-law was reported as the first observation of a slowing down of spreading~\cite{Mulansky_Ahnert_Pikovsky_11} and here we claim to have found a microscopic explanation of this effect.
The scaling of the chaos probability with system size for the harmonic oscillators in model B indicates that spreading states in this case are driven towards regular regions of the phase space.
There, they have to rely on thin chaotic layers which slows down their dynamics and leads to a decrease of the macroscopic spreading process.
In contrast, for model A the scaling of chaos probability predicts that spreading trajectories remain in the fully chaotic phase space regions which manifests in the macroscopic phenomenology of a pure power law spreading.

Note that in~\cite{Ivanchenko-Laptyeva-Flach-11} it was argued that in model B the probability to observe spreading of a state that is initially localized on $L$ sites with energy~$E$ converges to a finite non-zero value $\mathcal{P}_L>0$ in the limit of $L\rightarrow\infty$, i.e.\ ${W=E/L\rightarrow0}$.
Assuming that chaos and spreading are equivalent in this context, this would mean that the scaling exponent of chaos for model B is unity $\alpha=1$, i.e.\ the scaled variable in Figure~\ref{fig:2_4} should be $NW$.
This is a clear contradiction to the scaling with $\alpha=0.35$ reported above and the finding of our work is that the probability of chaos does approach zero in this limit.
In~\cite{basko2012local} it was argued that the authors of~\cite{Ivanchenko-Laptyeva-Flach-11} overestimated the spreading probability and the results presented here provide further evidence in this direction.
Indeed, considering that the limit ${W=E/L\rightarrow0}$ corresponds to $\beta\rightarrow0$ for the parametrization~\eqref{eqn:scaled_hamilton}, and thus the (integrable) limit of uncoupled harmonic oscillators, it is hard to imaging how a non-zero probability of spreading should be established.

\mysection{Conclusions}

In conclusion, we have analyzed the phase space structure of Hamiltonian chains of nonlinear or harmonic oscillators with nonlinear nearest neighbor coupling in terms of the probability of chaos~$\Pch$.
In extensive numerical simulations we were able to identify the scaling behavior of the transition point~$\Wtr$ with the system size~$N$ and found different scaling laws for nonlinear (model A) and harmonic (model B) oscillators.
Relating those scaling results to the spreading behavior studied widely in the past, we were able to explain the deviation from the pure power-law spreading reported recently.
Analyzing the properties of the microscopic chaos we were thus able to provide the first explanation for this slowing down of spreading, which has been controversially discussed in the recent past.
Although the comparison of spreading states with the properties of chaos for fixed sized chains seems physically reasonable, it should be noted that this is merely an assumption and requires further fortification.
Nevertheless, we believe that the results reported here provide new and valuable insight on how the properties of microscopic chaos influences macroscopic processes such as diffusion.

\mysection{Acknowledgements}

I thank A.~Pikovsky for numerous fruitful discussions.

\end{document}